\documentstyle [12pt]{article}
\begin{document}
\baselineskip 20.0 pt
\par
\mbox{}
\vskip -1.25in\par
\mbox{}
 \begin{flushright}
\makebox[1.5in][1]{UU-HEP-91/13}\\
\makebox[1.5in][1]{August 1991}
 \end{flushright}
 \vskip 0.25in

\begin{center}
{\large  \bf Bi-Hamiltonian Structure of Super KP Hierarchy}\\
\vspace{40 pt}
{Feng Yu }\\
\vspace{20 pt}
{{\it Department of Physics, University of Utah}}\\
{{\it Salt Lake City, Utah 84112, U.S.A.}}\\
\vspace{40 pt}
{\large ABSTRACT}\\
\end{center}
\vspace{10 pt}

We obtain the bi-Hamiltonian structure of the super KP hierarchy based on the
even super KP operator $\Lambda = \theta^{2} + \sum^{\infty}_{i=-2}U_{i}
\theta^{-i-1}$, as a supersymmetric extension of the ordinary KP
bi-Hamiltonian structure. It is expected to give rise to a universal super
$W$-algebra incorporating all known extended superconformal $W_{N}$ algebras
by reduction. We also construct the super BKP hierarchy by imposing a set of
anti-self-dual constraints on the super KP hierarchy.

\newpage
\section{Introduction}
\setcounter{equation}{0}
\vspace{5 pt}

The integrable nonlinear (partial) differential KdV [1] and KP [2] equations
are known to possess bi-Hamiltonian structures. This was first conjectured
by Adler [3] and then proved by Gelfand and Dickey [4] for the KdV hierarchy,
via the famous Gelfand-Dickey bracket. For the KP hierarchy, which is a set
of multi-time evolution equations in Lax form
\begin{eqnarray}
\frac{\partial L}{\partial t_{m}} = {[(L^{m})_{+} , L]}, ~~~~~m=1,2,3,\ldots~
\end{eqnarray}
based on the pseudo-differential operator
\begin{eqnarray}
L = D + \sum^{\infty}_{r=-1}u_{r}D^{-r-1},~~~~~D\equiv
\frac{\partial}{\partial x}
\end{eqnarray}
and contains all the $N$-th generalized KdV hierarchies by reduction
\begin{eqnarray}
(L^{m})_{-} = 0
\end{eqnarray}
where $(L^{m})_{+}, (L^{m})_{-}$ denote the purely differential and
pseudo-differential parts of $L^{m}$ respectively, the first Hamiltonian
structure was initially constructed by Watanabe [5] and the extension
to the bi-Hamiltonian structure was later worked out by Dickey [6].

There are two types of supersymmetric extensions of the KP hierarchy based on
the odd pseudo-super-differential operator [7]
\begin{eqnarray}
\bar{\Lambda} = \theta + \sum^{\infty}_{i=-1}\bar{U}_{i}
\theta^{-i-1}, ~~~~~\theta\equiv\frac{\partial}{\partial\zeta} + \zeta
\frac{\partial}{\partial x}
\end{eqnarray}
and the even pseudo-super-differential operator [8]
\begin{eqnarray}
\Lambda = \theta^{2} + \sum^{\infty}_{i=-2}U_{i}\theta^{-i-1}
\end{eqnarray}
respectively. (Note one can not obtain the odd (even) operator from the even
(odd) one.) In the former case, the discussion of its first Hamiltonian
structure [9] appeared to be pathological: Because of an inversion of the
even-odd parity of the nontrivial conserved supercharges (the integration of
the super Hamiltonian functions $SRes\bar{\Lambda}^{2n+1}$), neither these
supercharges nor the corresponding super Hamiltonian formalism can be
reduced to the ones of the ordinary KP hierarchy. To be of the
correct parity, one needs to start with the letter super hierarchy. It
precisely leads to the accomplishment of an explict constrction of the
first super KP Hamiltonian structure [10], which is indeed a supersymmetric
extension of the first ordinary KP Hamiltonian structure [5]. The reduction
of this super KP hierarchy gives rise to all the generalized $N$-th (even)
super KdV hierarchies which have been shown to be bi-Hamiltonian [8].
Thus it remains to explore the bi-Hamiltonian structure of the super
KP hierarchy.

Of physical interests, the KP hierarchy via its Hamiltonian structures is
known to incorporate all known extended conformal $W$-algebras. In particular,
the second Hamiltonian structure of the $N$-th KdV hierarchy was found [11]
to be identical to the $W_{N}$ algebra [12]; and then the first Hamiltonian
structure of the KP hierarchy itself was shown [13] to generate the
$W_{1+\infty}$ and $W_{\infty}$ algebras [14] that are the general linear
deformations of $W_{N}$ in the usual large $N$-limit [15]. Further, the
second KP Hamiltonian structure was suggested to be isomorphic to the unique
nonlinear deformation of  $W_{\infty}$ -- the $\hat{W}_{\infty}$ algebra
[16], which is expected to constitute a universal $W$-algebra containing
all $W_{N}$ algebras through the natural reduction of KP to KdV's.
Upon supersymmetrization, the newly established first super KP Hamiltonian
structure [10] naturally gives rise to a linear super $W_{1+\infty}$ algebra
whose bosonic sector is identified with $W_{1+\infty}\oplus W_{1+\infty}$,
and a subalgebra of it turns out to be isomorphic to the known $N=2$ super
$W_{\infty}$ algebra [17]. Similar to the ordinary case [16], one intends to
obtain a universal super $W$-algebra incorporating all known extended
superconformal $W_{N}$ algebras by reduction, and it is very likely to
be the general
nonlinear deformation of super $W_{\infty}$ which retains the characteristic
nonlinearity of super $W_{N}$. While there is evidence showing that super
$W_{N}$ are related to the super KdV hierarchies [18], this universal super
$W$-algebra is strongly expected to be the yet to be
determined second Hamiltonian structure of the super KP hierarchy.

The main issue of this paper is to obtain a one-parametric family of super
Hamiltonian forms based on the even operator (1.5), and identify the two
limiting cases of it as the first [10] and second Hamiltonian structures of
the corresponding super KP hierarchy. We will present a comprehensive proof
of this super bi-Hamiltonian structure in a way similar to the ordinary
KP case [6]. In addition, we will construct a super BKP hierarchy by imposing
a number of super anti-self-dual constraints on super KP, generalizing the
ordinary BKP hierarchy [19]. It may provide another class of super
$W$-algebras.

Since both KP (KdV) hierarchy and its Hamiltonian structures -- $W$-algebras
have played essential roles in 2d quantum gravity and noncritical string
theories [20,21,22], the determination of the super KP bi-Hamiltomian
structure will naturally provide a promising framework of studying 2d
quantum supergravity and noncritical superstrings.

\vspace{30 pt}
\section{Super KP Hierarchy in Hamiltonian Formalism}
\setcounter{equation}{0}
\vspace{5 pt}

We will formulate the super KP hierarchy on $(1\mid 1)$ superspace. The basic
variables are $x$ and $\zeta$ with parity even and odd respectively. The
supercovariant derivative $\theta = \partial/\partial\zeta +
\zeta\partial/\partial x$ satisfies $\theta^{2}=\partial(\equiv\partial
/\partial x)$. The basic ingredient of super KP hierarchy is a
pseudo-super-differential operator with even parity given by eq.(1.5), where
the superfield coefficients $U_{i}$ are functions of $x,\zeta$ and
various (even) time variables $t_{m} (m=1,2,3, \ldots)$. The parity of a
function $F$ will be indicated by $p(F)$ which is equal to zero for $F$ being
even and one for $F$ being odd. Accordingly, $p(U_{i})=i+1$. Recall that an
arbitrary pseudo-super-differential operator $P$ has the formal expression
\begin{eqnarray}
P = \sum_{i=-\infty}^{N}V_{i}\theta^{i}
\end{eqnarray}
and
\begin{eqnarray}
P_{+} \equiv \sum_{i=0}^{N}V_{i}\theta^{i},~~~~~P_{-} \equiv
\sum_{i=-\infty}^{-1}V_{i}\theta^{i}.
\end{eqnarray}
The multiplication of two such operators $P$ and $Q$
is determined by the associativity and the basic relation $\theta UV =
(\theta U)V+(-1)^{p(U)}U\theta V$ for $\theta$ acting on arbitrary superfield
functions $U$ and $V$; in particular,
\begin{eqnarray}
\theta^{i}U = \sum^{\infty}_{l=0}(-1)^{p(U)(i-l)}\left[ \begin{array}{c}
i\\l
\end{array} \right] U^{[l]}\theta^{i-l}
\end{eqnarray}
where $U^{[l]}\equiv(\theta^{l}U)$ that is different from $\theta^{l}U$. The
super-binomial coefficients $\left[ \begin{array}{c}
i\\k
\end{array} \right]$ turn out to be for $i\geq 0$
\begin{eqnarray}
\left[ \begin{array}{c}
i\\k
\end{array} \right] = \left\{ \begin{array}{ll}
0     & \mbox{for $k<0$ or $k>i$ or $(i,k)=(0,1)$ mod $2$} \\
\left( \begin{array}{c}
{[i/2]} \\ {[k/2]}
\end{array} \right)  & \mbox{for $0\leq k\leq i$ and $(i,k)\neq(0,1)$ mod $2$}
\end{array} \right. ;
\end{eqnarray}
and are expressed for $i<0$ by the identity
\begin{eqnarray}
\left[ \begin{array}{c}
i\\k
\end{array} \right] = (-1)^{[k/2]}\left[ \begin{array}{c}
-i+k-1\\k
\end{array} \right].
\end{eqnarray}
The super KP hierarchy in the Lax form is a system of infinitely many evolution
equations for the functions $U_{i}$
\begin{eqnarray}
\frac{\partial\Lambda}{\partial t_{m}} = {[(\Lambda^{m})_{+} , \Lambda]}.
\end{eqnarray}
It is easy to check that the different time evolutions in eq.(2.6)
are consistent, for these flows actually commute with each other:
\begin{eqnarray}
\frac{\partial^{2}\Lambda}{\partial t_{m}\partial t_{n}} =
\frac{\partial^{2}\Lambda}{\partial t_{n}\partial t_{m}}.
\end{eqnarray}
This super KP hierarchy has four times as many
degrees of freedom as its bosonic
counterpart has. By letting
$U_{2r+1}=v_{2r}=0$ $(U_{i}(x,\zeta)\equiv$ $v_{i}(x)+
\zeta u_{i}(x))$, it reduces, though not manifestly, indeed to the ordinary
KP hierarchy [2]. Particularly, it owns infinitely many independent conserved
supercharges $\int\int(1/n)SRes\Lambda^{n}dxd\zeta$ $\equiv\int(1/n)
SRes\Lambda^{n}dX$:
\begin{eqnarray}
\frac{\partial}{\partial t_{m}}\int\frac{1}{n}SRes\Lambda^{n} dX = 0~~~~~~
n=1,2,3,\ldots~,
\end{eqnarray}
which are truly the supersymmetric extension of the conserved charges in the
KP hierarchy. Here the super-residue of a super operator $P$,
$SResP$, means the coefficient of $\theta^{-1}$ term in $P$. The proof of
eq.(2.8) is indebted to the powerful theorem on the super commutator of
two pseudo-super-differential operators $P$ and $Q$ [7]:
\begin{eqnarray}
\int SRes{[P , Q]}dX = 0,
\end{eqnarray}
where $[P,Q]$ is defined to be $PQ-(-1)^{p(P)p(Q)}QP$.

In searching for the super KP Hamiltonian structures, one tries to put
eq.(2.6) into the Hamiltonian form
\begin{eqnarray}
\frac{\partial\Lambda}{\partial t_{m}} = K\frac{\delta\Pi_{m}}{\delta U}
= \sum^{\infty}_{i,j=-2}(-1)^{j}K_{ij}
\frac{\delta\Pi_{m}}{\delta U_{j}}\theta^{-i-1},
\end{eqnarray}
where the factor $(-1)^{j}$ is introduced to maintain the correct parity and
$\delta/\delta U_{i}$ stands for the variational derivative for superfield
functions
\begin{eqnarray}
\frac{\delta F}{\delta U_{i}} = \sum_{k=0}(-1)^{(i+1)k+k(k+1)/2}
(\frac{\partial F}{\partial U_{i}^{[k]}})^{[k]}.
\end{eqnarray}
The infinite dimensional supermatrix $K_{ij}$ in eq.(2.10)
is said to be a Hamiltonian
structure if the super Poisson brackets associated with it
\begin{eqnarray}
{\{U_{i}(X) , U_{j}(Y)\}} = K_{ij}(X)\delta(X-Y)
\end{eqnarray}
form an algebra, that is equivalent to proving the brackets between
arbitrary superfield functions $F(U_{i})$ and $G(U_{j})$
\begin{eqnarray}
& & {\{\int F(U_{i}(X))dX , \int G(U_{j}(Y))dY\}} \nonumber\\
&=& \sum_{i,j=-2}^{\infty}
\int (-1)^{(p(F)+1)(i+1)}\frac{\delta F(X)}{\delta U_{i}(Z)}K_{ij}(Z)
\frac{\delta G(Y)}{\delta U_{j}(Z)} dZ
\end{eqnarray}
satisfy the super Jacobi identities and are super-antisymmetric. Here
$X\equiv (x,\zeta_{x})$ and $\delta(X-Y)\equiv\delta(x-y)\delta(\zeta_{x}
-\zeta_{y})$. Correspondingly, $\Pi_{m}$ in eq.(2.10) are regarded as super
Hamiltonian functions of this Hamiltonian structure. Now with
respect to eq.(2.12), one is able to rewrite the Hamiltonian form (2.10)
into algebraic brackets
\begin{eqnarray}
\frac{\partial U_{i}}{\partial t_{m}} = {[(\Lambda^{m})_{+} , \Lambda]}_{i}
={\{U_{i} , \int \Pi_{m}(Y)dY\}}
\end{eqnarray}
which will appear to be a convenient form in identifying the Hamiltonian
structures later.

To proceed, let us expand $\Lambda^{m}$ as
\begin{eqnarray}
\Lambda^{m} = \sum^{2m}_{j=-\infty}\theta^{j}\lambda_{j}(m).
\end{eqnarray}
It follows that [10]
\begin{eqnarray}
\lambda_{j}(m) = (-1)^{j}\frac{1}{(m+1)}\frac{\delta SRes\Lambda^{m+1}}
{\delta U_{j}}~~~~~~j\geq 0.
\end{eqnarray}
Then use the above expression to calculate
${[(\Lambda^{m})_{+} , \Lambda]}$. This leads directly to the first
Hamiltonian form of eq.(2.6)
\begin{eqnarray}
{[(\Lambda^{m})_{+} , \Lambda]}_{i} &=& \frac{1}{(m+1)}\sum^{\infty}_{j=-2}
(-1)^{j}K_{ij}^{(1)}\frac{\delta SRes\Lambda^{m+1}}{\delta U_{j}} \nonumber\\
&=& {\{U_{i} , \frac{1}{m+1}\int SRes\Lambda^{m+1}(Y)dY\}}_{1}
\end{eqnarray}
where $K^{(1)}_{ij}$ (the index 1 denotes the first one, similar for 2 below)
has been proved to be the first Hamiltonian structure in ref.[10] and the
conserved supercharge densities $(1/m+1)SRes\Lambda^{m+1}$ are identified with
the super Hamiltonian functions $\Pi_{m}^{(1)}$ as expected.

To obtain the second Hamiltonian form of eq.(2.6), let us express
${[(\Lambda^{m})_{+},\Lambda]}$ into a bilinear form of $\Lambda$ (similar
to the ordinary KP case)
\begin{eqnarray}
{[(\Lambda^{m})_{+},\Lambda]} = (\Lambda\Lambda^{m-1})_{+}\Lambda-
\Lambda(\Lambda^{m-1}\Lambda)_{+}.
\end{eqnarray}
Then we substitute eqs.(2.15)-(2.16) for $\Lambda^{m-1}$ into the right hand
side of eq.(2.18), and it turns out to be
\begin{eqnarray}
((\Lambda\Lambda^{m-1})_{+}\Lambda-\Lambda(\Lambda^{m-1}\Lambda)_{+})_{i}
&=& \frac{1}{m}\sum^{\infty}_{j=-2}
(-1)^{j}K_{ij}^{(2)}\frac{\delta SRes\Lambda^{m}}{\delta U_{j}} \nonumber\\
&=& {\{U_{i} , \frac{1}{m}\int SRes\Lambda^{m}(Y)dY\}}_{2}
\end{eqnarray}
where $K^{(2)}_{ij}$ is the yet to be determined second super KP Hamiltonian
structure with $(1/m)SRes\Lambda^{m}$ begin the associated super Hamiltonian
functions $\Pi_{m}^{(2)}$. We leave the proof of this second Hamiltonian
structure to the next section.

\vspace{30 pt}
\section{Bi-Hamiltonian Structure}
\setcounter{equation}{0}
\vspace{5 pt}

We observe from eq.(2.19) that to obtain the second Hamiltonian structure of
eq.(2.6) it is crucial to analyze the mapping from the set of all
pseudo-super-differential operators
\begin{eqnarray}
R \equiv \{M = \sum_{i}\theta^{i}m_{i}\}
\end{eqnarray}
to a special set of operators $S\equiv\{\sum^{\infty}_{i=-2}
V_{i}\theta^{-i-1}\}$ (note $\Lambda \in S$):
\begin{eqnarray}
K(M) = (\Lambda M)_{+}\Lambda - \Lambda(M\Lambda)_{+} = \Lambda(M\Lambda)_{-}
-(\Lambda M)_{-}\Lambda.
\end{eqnarray}
By a constant shift of $\Lambda$ to $\hat{\Lambda}\equiv\Lambda -c$, the
mapping
\begin{eqnarray}
\hat{K}(M) = (\hat{\Lambda}M)_{+}\hat{\Lambda} - \hat{\Lambda}
(M\hat{\Lambda})_{+} = \sum^{\infty}_{i=-2}K_{i}(M)\theta^{-i-1}
\end{eqnarray}
will actually give rise to a one-parametric family of super KP Hamiltonian
forms; by taking $c=0$, eq.(3.3) obviously goes back to eq.(3.2) and in the
$c\rightarrow\infty$ limit, eq.(3.3) is reduced to the first Hamiltonian form
with $M=\Lambda^{n}$:
\begin{eqnarray}
\lim_{c\rightarrow\infty}(-\frac{1}{c}\hat{K}(M)) &=& M_{+}\Lambda +
(\Lambda M)_{+} - (M\Lambda)_{+}- \Lambda M_{+} \nonumber\\
&=& {[M_{+},\Lambda]} +{[\Lambda,M]} ~=~ {[\Lambda^{n}_{+},\Lambda]}.
\end{eqnarray}
Thus in general, one may consider the bi-Hamiltonian mapping (3.3).

To begin with, let us assign a super-differentiation $\theta_{\alpha}$ to
each pseudo-super-differential operator $\alpha = \sum^{\infty}_{i=-2}
\alpha_{i}\theta^{-i-1} \in S$. It acts on superfield fuction $F$ as
\begin{eqnarray}
\theta_{\alpha}F = \sum^{\infty}_{i=0}\sum^{\infty}_{j=-2}(-1)^{p(\alpha)i}
\alpha_{j}^{[i]}\frac{\partial F}{\partial U_{j}^{[i]}},
\end{eqnarray}
and the action can directly extend onto an arbitrary pseudo-super-differential
operator $P =\sum_{i}V_{i}\theta^{i}$ as
\begin{eqnarray}
\theta_{\alpha}P = \sum_{i}(\theta_{\alpha}V_{i})\theta^{i}.
\end{eqnarray}
Nontrivially, this super-differentiation supercommutes with $\theta$:
\begin{eqnarray}
{[\theta_{\alpha} , \theta]} =0
\end{eqnarray}
and therefore can be well-defined on the super-functionals $\int F(U_{i}(X))
dX$ under appropriate boundary conditions as
\begin{eqnarray}
\theta_{\alpha}\int F dX = \int \theta_{\alpha}F dX.
\end{eqnarray}
Now we are going to prove the following proposition:
\begin{eqnarray}
{[\theta_{\hat{K}(M)},\theta_{\hat{K}(N)}]} =
\theta_{\hat{K}(M(\hat{\Lambda}N)_{-}-(M\hat{\Lambda})_{+}N
+\theta_{\hat{K}(M)}N-(-1)^{p(M)p(N)}(M\leftrightarrow N))}.
\end{eqnarray}

Proof: First we expand the right hand side of eq.(3.9) on superfield $F$:
\begin{eqnarray}
& & {[\theta_{\hat{K}(M)},\theta_{\hat{K}(N)}]}F \nonumber\\
&=& \sum_{i,i'=0}\sum_{j,j'=-2}(-1)^{p(N)i+p(M)i'}K_{j'}^{[i']}(M)
\frac{\partial (K_{j}^{[i]}(N)\frac{\partial F}{\partial
U_{j}^{[i]}})}{\partial U_{j'}^{[i']}} -(-1)^{p(M)p(N)}(M\leftrightarrow N)
\nonumber\\
&=& \sum_{i,i'=0}\sum_{j,j'=-2}(-1)^{p(N)i+p(M)i'}K_{j'}^{[i']}(M)
\frac{\partial K_{j}^{[i]}(N)}{\partial U_{j'}^{[i']}}
\frac{\partial F}{\partial U_{j}^{[i]}}
-(-1)^{p(M)p(N)}(M\leftrightarrow N) \nonumber\\
&=& \sum_{i=0}\sum_{j=-2}(-1)^{(p(M)+p(N))i} (\theta_{\hat{K}(M)}
K_{j}(N))^{[i]}\frac{\partial F}{\partial U_{j}^{[i]}}
-(-1)^{p(M)p(N)}(M\leftrightarrow N) \nonumber\\
&=& \theta_{\theta_{\hat{K}(M)}\hat{K}(N) -(-1)^{p(M)p(N)}\theta_{\hat{K}(N)}
\hat{K}(M)} F
\end{eqnarray}
where we have used eq.(3.7) and have taken into account the
$(-1)^{p(M)p(N)}(M\leftrightarrow N)$ part. Next one only needs to evaluate
\begin{eqnarray}
A &\equiv& \theta_{\hat{K}(M)}\hat{K}(N) -(-1)^{p(M)p(N)}\theta_{\hat{K}(N)}
\hat{K}(M) \nonumber\\
&=& \theta_{\hat{K}(M)}((\hat{\Lambda}N)_{+}\hat{\Lambda}-\hat{\Lambda}
(N\hat{\Lambda})_{+}) -(-1)^{p(M)p(N)}(M\leftrightarrow N). \nonumber
\end{eqnarray}
Noticing that $\theta_{\alpha}\Lambda = \alpha$  implied by eq.(3.6) and
using eq.(3.7) again, we have
\begin{eqnarray}
A &=& (\hat{K}(M)N)_{+}\hat{\Lambda}+(\hat{\Lambda}\theta_{\hat{K}(M)}N)_{+}
\hat{\Lambda} +(-1)^{p(M)p(N)}(\hat{\Lambda}N)_{+}\hat{K}(M) -\hat{K}(M)
(N\hat{\Lambda})_{+} \nonumber\\
& & -\hat{\Lambda}((\theta_{\hat{K}(M)}N)\hat{\Lambda})_{+}
-(-1)^{p(M)p(N)}\hat{\Lambda}(N\hat{K}(M))_{+} -(-1)^{p(M)p(N)}
(M\leftrightarrow N) \nonumber\\
&=& (((\hat{\Lambda}M)_{+}\hat{\Lambda} -\hat{\Lambda}(M\hat{\Lambda})_{+})
N)_{+}\hat{\Lambda}+(-1)^{p(M)p(N)}(\hat{\Lambda}N)_{+}
((\hat{\Lambda}M)_{+}\hat{\Lambda} -\hat{\Lambda}(M\hat{\Lambda})_{+})
\nonumber\\
& & -((\hat{\Lambda}M)_{+}\hat{\Lambda} -\hat{\Lambda}(M\hat{\Lambda})_{+})
(N\hat{\Lambda})_{+} -(-1)^{p(M)p(N)}\hat{\Lambda}
(N((\hat{\Lambda}M)_{+}\hat{\Lambda} -\hat{\Lambda}(M\hat{\Lambda})_{+}))_{+}
\nonumber\\
& & +\hat{K}(\theta_{\hat{K}(M)}N) -(-1)^{p(M)p(N)}(M\leftrightarrow N).
\nonumber
\end{eqnarray}
It follows by taking the $(-1)^{p(M)p(N)}(M\leftrightarrow N)$ terms into
account more frequently that
\begin{eqnarray}
A &=& ((\hat{\Lambda}M)_{+}\hat{\Lambda}N -\hat{\Lambda}(M\hat{\Lambda})_{+}
N)_{+}\hat{\Lambda}+(-1)^{p(M)p(N)}(\hat{\Lambda}N)_{+}
(\hat{\Lambda}M)_{+}\hat{\Lambda} \nonumber\\
& & +\hat{\Lambda}(M\hat{\Lambda})_{+}
(N\hat{\Lambda})_{+}
-(-1)^{p(M)p(N)}\hat{\Lambda}
(N(\hat{\Lambda}M)_{+}\hat{\Lambda} -N\hat{\Lambda}(M\hat{\Lambda})_{+})_{+}
\nonumber\\
& &
+\hat{K}(\theta_{\hat{K}(M)}N) -(-1)^{p(M)p(N)}(M\leftrightarrow N) \nonumber\\
&=& ((\hat{\Lambda}M)_{+}\hat{\Lambda}N -\hat{\Lambda}(M\hat{\Lambda})_{+}N
-(\hat{\Lambda}M)_{+}(\hat{\Lambda}N)_{+})_{+}\hat{\Lambda}
+\hat{\Lambda}((M\hat{\Lambda})_{+}(N\hat{\Lambda})_{+} \nonumber\\
& & +M(\hat{\Lambda}N)_{+}\hat{\Lambda} -M\hat{\Lambda}
(N\hat{\Lambda})_{+})_{+}
+\hat{K}(\theta_{\hat{K}(M)}N) -(-1)^{p(M)p(N)}(M\leftrightarrow N) \nonumber\\
&=& (\hat{\Lambda}M(\hat{\Lambda}N)_{-} -\hat{\Lambda}(M\hat{\Lambda})_{+}
N)_{+}\hat{\Lambda} -\hat{\Lambda}((M\hat{\Lambda})_{-}N\hat{\Lambda}
-M(\hat{\Lambda}N)_{+}\hat{\Lambda})_{+} \nonumber\\
& & +\hat{K}(\theta_{\hat{K}(M)}N) -(-1)^{p(M)p(N)}(M\leftrightarrow N)
\nonumber\\
&=& \hat{K}(M(\hat{\Lambda}N)_{-} -(M\hat{\Lambda})_{+}N)
 +\hat{K}(\theta_{\hat{K}(M)}N) -(-1)^{p(M)p(N)}(M\leftrightarrow N)
\nonumber\\
&=& \hat{K}(M(\hat{\Lambda}N)_{-} -(M\hat{\Lambda})_{+}N
+\theta_{\hat{K}(M)}N -(-1)^{p(M)p(N)}(M\leftrightarrow N))
\end{eqnarray}
as desired. (QED)

This proposition allows us to define a closed 2-form $\Omega$ on the set of
super-differentiations $\{\theta_{\hat{K}(M)}\}$:
\begin{eqnarray}
\Omega (\theta_{\hat{K}(M)},\theta_{\hat{K}(N)}) \equiv
\int SRes(\hat{K}(M)N)dX,
\end{eqnarray}
which is necessarily super-antisymmetric:
\begin{eqnarray}
\Omega (\theta_{\hat{K}(N)},\theta_{\hat{K}(M)})
= -(-1)^{p(M)p(N)}\Omega (\theta_{\hat{K}(M)},\theta_{\hat{K}(N)}).
\end{eqnarray}
The closeness of eq.(3.12) constitutes another proposition.

Proof: With a suitable parity factor in front of each term, we have
\begin{eqnarray}
& & d\Omega (\theta_{\hat{K}(M)},\theta_{\hat{K}(N)}, \theta_{\hat{K}(L)})
\nonumber\\
&=& (-1)^{p(M)p(L)}(\theta_{\hat{K}(M)}\Omega (\theta_{\hat{K}(N)},
\theta_{\hat{K}(L)}) -\Omega ({[\theta_{\hat{K}(M)},\theta_{\hat{K}(N)}]}
,\theta_{\hat{K}(L)})) + c.p.
\end{eqnarray}
where $c.p.$ is understood as the overall cyclic permutations among $M,N,L$.
The first term of eq.(3.14) is well-defined due to eq.(3.8) and is evaluated
as follows:
\begin{eqnarray}
& & \theta_{\hat{K}(M)}\Omega(\theta_{\hat{K}(N)},\theta_{\hat{K}(L)})
\nonumber\\
&=& \int SRes(\theta_{\hat{K}(M)}((\hat{\Lambda}N)_{+}\hat{\Lambda}
-\hat{\Lambda}(N\hat{\Lambda})_{+})L)dX \nonumber\\
&=& \int SRes{\{ ((\hat{K}(M)N)_{+}\hat{\Lambda} +(-1)^{p(M)p(N)}
(\hat{\Lambda}N)_{+}\hat{K}(M) -\hat{K}(M)(N\hat{\Lambda})_{+} }\nonumber\\
& & ~~~{-(-1)^{p(M)p(N)}\hat{\Lambda}(N\hat{K}(M))_{+})L +\hat{K}
(\theta_{\hat{K}(M)}N)L +(-1)^{p(M)p(N)}\hat{K}(N)\theta_{\hat{K}(M)}L \}}dX
\nonumber\\
&=& \int SRes{\{ \hat{K}(M)(N(\hat{\Lambda}L)_{-} -(N\hat{\Lambda})_{+}L
-(-1)^{p(N)p(L)}(N\leftrightarrow L)) } \nonumber\\
& & ~~~{-(-1)^{p(L)(p(M)+p(N))}\hat{K}(L)
\theta_{\hat{K}(M)}N +(-1)^{p(M)p(N)}\hat{K}(N)\theta_{\hat{K}(M)}L \}}dX.
\end{eqnarray}
The evaluation of the second term of eq.(3.14) needs the proposition (3.9)
and it turns out to be
\begin{eqnarray}
& & -\Omega ({[\theta_{\hat{K}(M)},\theta_{\hat{K}(N)}]},\theta_{\hat{K}(L)})
\nonumber\\
&=& \int SRes{\{ (-1)^{p(L)(p(M)+p(N))}
\hat{K}(L)(M(\hat{\Lambda}N)_{-} -(M\hat{\Lambda})_{+}N } \nonumber\\
& & ~~~{+\theta_{\hat{K}(M)}N  -(-1)^{p(M)p(N)}(M\leftrightarrow N)) \}}dX .
\end{eqnarray}
Now substituting eqs.(3.15)-(3.16) into (3.14), we find
\begin{eqnarray}
& & d\Omega (\theta_{\hat{K}(M)},\theta_{\hat{K}(N)}, \theta_{\hat{K}(L)})
\nonumber\\
&=& \int SRes{\{ ((-1)^{p(M)p(L)}
\hat{K}(M)(N(\hat{\Lambda}L)_{-} -(N\hat{\Lambda})_{+}L
-(-1)^{p(N)p(L)}(N\leftrightarrow L)) } \nonumber\\
& & ~~~{+ (M\leftrightarrow L, N\leftrightarrow M,
L\leftrightarrow N)) } \nonumber\\
& & ~~~{ +(-1)^{p(M)(p(N)+p(L))}\hat{K}(N)
\theta_{\hat{K}(M)}L -(-1)^{p(N)(p(L)+p(M))}\hat{K}(L)
\theta_{\hat{K}(N)}M \}}dX + c.p. \nonumber\\
&=& 2\int SRes{\{ (-1)^{p(M)p(L)}
((\hat{\Lambda}M)_{+}\hat{\Lambda}-\hat{\Lambda}(M\hat{\Lambda})_{+})
(N(\hat{\Lambda}L)_{-} -(N\hat{\Lambda})_{+}L }\nonumber\\
& & ~~~{-(-1)^{p(N)p(L)}(N\leftrightarrow L)) \} } dX +c.p.. \nonumber
\end{eqnarray}
Further, by using eq.(2.9) and the following lemma:
\begin{eqnarray}
(-1)^{p(P)p(R)} \int SRes(P_{-}QR_{+})dX + c.p. = (-1)^{p(P)p(R)}\int SRes(PQR)
dX,
\end{eqnarray}
we conclude that
\begin{eqnarray}
& & d\Omega (\theta_{\hat{K}(M)},\theta_{\hat{K}(N)}, \theta_{\hat{K}(L)})
\nonumber\\
&=& 2(-1)^{p(M)p(L)} \int SRes
((\hat{\Lambda}M)_{-}\hat{\Lambda}N(\hat{\Lambda}L)_{+}-
(M\hat{\Lambda})_{+}(N\hat{\Lambda})_{-}L\hat{\Lambda} \nonumber\\
& & ~~~-(-1)^{p(N)p(L)}(N\leftrightarrow L)) dX +c.p. \nonumber\\
&=& 2(-1)^{p(M)p(L)} \int SRes
(\hat{\Lambda}M\hat{\Lambda}N\hat{\Lambda}L-
M\hat{\Lambda}N\hat{\Lambda}L\hat{\Lambda}
-(-1)^{p(N)p(L)}(N\leftrightarrow L)) dX  \nonumber\\
&=& 0.
\end{eqnarray}
(QED)

Before we define the super Poisson brackets associated with this closed
2-form, let us introduce the variational derivative with respect to the super
KP operator $\Lambda$ as an generalization of eq.(2.11):
\begin{eqnarray}
\frac{\delta F}{\delta \Lambda} = \sum^{\infty}_{i=-2}(-1)^{p(F)+i+1}
\theta^{i}\frac{\delta F}{\delta U_{i}}.
\end{eqnarray}
It follows that
\begin{eqnarray}
\theta_{\alpha}\int F dX = \int SRes(\alpha\frac{\delta F}{\delta\Lambda})dX.
\end{eqnarray}
Comparing eq.(3.20) with (3.12), one finds eq.(3.12) with $N$ chosen to be
$\delta F/\delta\Lambda$ may be expressed in terms of the
super-differentiation $\theta_{\hat{K}(M)}$ on the functional $\int FdX$:
\begin{eqnarray}
\Omega (\theta_{\hat{K}(M)},\theta_{\hat{K}(\frac{\delta F}{\delta\Lambda})})
= \int SRes (\hat{K}(M)\frac{\delta F}{\delta\Lambda})dX =\theta_{\hat{K}(M)}
\int FdX.
\end{eqnarray}
Note eqs.(3.20)-(3.21) are well-defined due to eq.(3.8) so that we can
denote $\theta_{\hat{K}(\delta F/\delta\Lambda)}$ as
$\theta_{\int FdX}$. Now taking $M$ to be $\delta G/\delta\Lambda$, we are
led to the following definition of super Poisson brackets between the
functionals $\int FdX$ and $\int GdX$:
\begin{eqnarray}
{\{\int F(X)dX, \int G(Y)dY\}} =\Omega (\theta_{\int FdX},\theta_{\int GdY})
=\theta_{\int FdX}\int GdY.
\end{eqnarray}
Implied by eqs.(3.13), eq.(3.22) is super-antisymmetric as required.
Furthermore, with eq.(3.21), one can easily show the next proposition:
\begin{eqnarray}
\theta_{\{\int FdX, \int GdY\}} = {[\theta_{\int FdX},\theta_{\int GdY}]}.
\end{eqnarray}

Proof: It is equivalent to show
\begin{eqnarray}
I\equiv \Omega(\theta_{\hat{K}(L)}, -{[\theta_{\int FdX},\theta_{\int GdY}]}
+ \theta_{\{\int FdX, \int GdY\}}) =  0
\end{eqnarray}
for arbitrary $L$. Indeed, by adding some proper null terms,
\begin{eqnarray}
I &=& \theta_{\hat{K}(L)}{\{\int FdX, \int GdY\}} -
\Omega(\theta_{\hat{K}(L)}, {[\theta_{\int FdX},\theta_{\int GdY}]})
\nonumber\\
&=& -(-1)^{(p(F)+1)(p(G)+1)}\theta_{\hat{K}(L)}\theta_{\int GdY}\int FdX
-\theta_{\hat{K}(L)}\theta_{\int FdX}\int GdY \nonumber\\
& & + (-1)^{p(L)(p(F)+1)}
\theta_{\int FdX}\theta_{\hat{K}(L)}\int GdY +
{[\theta_{\hat{K}(L)},\theta_{\int FdX}]}\int GdY \nonumber\\
& & +(-1)^{p(L)(p(F)+p(G)+2)}
\Omega({[\theta_{\int FdX},\theta_{\int GdY}]},\theta_{\hat{K}(L)}) \nonumber\\
&=& -(-1)^{(p(G)+1)(p(F)+p(L)+1)}\theta_{\int GdY}\theta_{\hat{K}(L)}\int FdX
\nonumber\\
& & - (-1)^{(p(F)+1)(p(G)+1)}{[\theta_{\hat{K}(L)},\theta_{\int GdY}]}\int FdX
\nonumber\\
& & -\theta_{\hat{K}(L)}\theta_{\int FdX}\int GdY + (-1)^{p(L)(p(F)+1)}
\theta_{\int FdX}\theta_{\hat{K}(L)}\int GdY \nonumber\\
& & +{[\theta_{\hat{K}(L)},\theta_{\int FdX}]}
\int GdY +(-1)^{p(L)(p(F)+p(G)+2)}
\Omega({[\theta_{\int FdX},\theta_{\int GdY}]},\theta_{\hat{K}(L)}) \nonumber\\
&=& -(\theta_{\hat{K}(L)}\Omega(\theta_{\int FdX},\theta_{\int GdY})
-\Omega({[\theta_{\hat{K}(L)},\theta_{\int FdX}]},\theta_{\int GdY})
\nonumber\\
& & -(-1)^{p(L)(p(F)+p(G)+2)}(\theta_{\int FdX}\Omega(\theta_{\int GdY},
\theta_{\hat{K}(L)}) -\Omega({[\theta_{\int FdX},\theta_{\int GdY}]},
\theta_{\hat{K}(L)})) \nonumber\\
& & -(-1)^{(p(G)+1)(p(F)+p(L)+1)}(\theta_{\int GdY}\Omega(\theta_{\hat{K}(L)},
\theta_{\int FdX}) -\Omega({[\theta_{\int GdY},\theta_{\hat{K}(L)}]},
\theta_{\int FdX})) \nonumber\\
&=& (-1)^{p(L)(p(G)+1)+1}d\Omega(\theta_{\hat{K}(L)},\theta_{\int FdX},
\theta_{\int GdY})
\end{eqnarray}
which vanishes because of eq.(3.18). (QED)

As a consequence, the super Poisson
brackets (3.22) satisfy the super Jacobi identities. Theorem:
\begin{eqnarray}
J \equiv (-1)^{(p(F)+1)(p(H)+1)}{\{ {\{\int FdX,\int GdY\}}, \int HdZ \}}
+ c.p. = 0.
\end{eqnarray}

Proof: On one hand,
\begin{eqnarray}
J &=& -(-1)^{(p(F)+1)(p(H)+1)}{\{ \int FdX, {\{\int GdY, \int HdZ\}} \}}
+ c.p. \nonumber\\
&=&  -(-1)^{(p(F)+1)(p(H)+1)}\theta_{\int FdX}{\{\int GdY, \int HdZ\}}
+ c.p. \nonumber\\
&=&  -(-1)^{(p(F)+1)(p(H)+1)}\theta_{\int FdX}\Omega(\theta_{\int GdY},
\theta_{\int HdZ}) + c.p. , \nonumber
\end{eqnarray}
on the other hand, by using eq.(3.25),
\begin{eqnarray}
J &=& (-1)^{(p(F)+1)(p(H)+1)}\theta_{\{\int FdX,\int GdY\}}\int HdZ
+ c.p. \nonumber\\
&=& (-1)^{(p(F)+1)(p(H)+1)}\Omega(\theta_{\{\int FdX,\int GdY\}},
\theta_{\int HdZ}) + c.p. . \nonumber
\end{eqnarray}
Thus,
\begin{eqnarray}
2J = -d\Omega(\theta_{\int FdX},\theta_{\int GdY},\theta_{\int HdZ}) = 0.
\end{eqnarray}
(QED)

So we have proved the super Poisson algebra (3.22) based on the general
Hamiltonian form (3.3) for arbitrary pseudo-super-differential operator $M$.
It remains to specify $M$ so that one may use this superalgebra to indeed
obtain the bi-Hamiltonian structure of the super KP hierarchy. Let us first
set $M=\Lambda^{m-1}=\sum_{j}\theta^{j}\lambda_{j}(m-1)$
and the parameter $c=0$. Note only the $j\geq -2$ part of $M$ is involved in
eq.(3.3); we denote it as $\Lambda^{m-1}_{\tilde{+}}$. From eq.(2.16), we
have
\begin{eqnarray}
\Lambda^{m-1}_{\tilde{+}} &=& \sum^{\infty}_{j=-2}\theta^{j}\lambda_{j}(m-1)
\frac{\delta SRes\Lambda^{m}}{\delta U_{j}} \nonumber\\
&=& \frac{1}{m}\frac{\delta SRes\Lambda^{m}}{\delta \Lambda}.
\end{eqnarray}
Hence, we have corrspondingly chosen the function $F$ in eq.(3.22) to be
the super Hamiltonian function $\Pi_{m}^{(2)} = (1/m)SRes\Lambda^{m}$. Then,
\begin{eqnarray}
& & {\{ \frac{1}{m}\int SRes\Lambda^{m}dX, \int GdY\}}_{2} \nonumber\\
&=&
\theta_{(1/m)\int SRes\Lambda^{m}dX}\int GdY ~=~ \int SRes(K(\Lambda^{m-1})
\frac{\delta G}{\delta\Lambda})dX \nonumber\\
&=& \int SRes({[(\Lambda^{m})_{+},\Lambda]}\frac{\delta G}{\delta\Lambda})dX
\end{eqnarray}
which is equivalent to
\begin{eqnarray}
{\{ \frac{1}{m}\int SRes\Lambda^{m}dX, U_{i} \}}_{2}
= -{\{ U_{i}, \int\Pi^{(2)}_{m}dX \}}_{2} = {[(\Lambda^{m})_{+},\Lambda]}_{i}.
\end{eqnarray}
Except for an insignificant negative sign (which can be absorbed into a
redefinition of eq.(3.22)), eq.(3.30) precisely coincides with the
Hamiltonian form (2.19) and thus gives rise to the second Hamiltonian
structure of the super KP hierarchy (2.6).

Now we set $M=\Lambda^{m}$ and let $c\rightarrow\infty$. This amounts to choose
$F$ to be $\Pi^{(1)}_{m}=(1/(m+1))SRes\Lambda^{m+1}$ and then
\begin{eqnarray}
& & {\{ \frac{1}{m+1}\int SRes\Lambda^{m+1}dX, \int GdY\}}_{1} \nonumber\\
&=&
\lim_{c\rightarrow\infty}\frac{1}{c}\int SRes(\hat{K}(\Lambda^{m})
\frac{\delta G}{\delta\Lambda})dX ~=~
\int SRes({[\Lambda, (\Lambda^{m})_{+}]}\frac{\delta G}{\delta\Lambda})dX
\end{eqnarray}
or
\begin{eqnarray}
-{\{ \frac{1}{m+1}\int SRes\Lambda^{m+1}dX, U_{i} \}}_{1}
= {\{ U_{i}, \int\Pi^{(1)}_{m}dX \}}_{1} = {[(\Lambda^{m})_{+},\Lambda]}_{i}.
\end{eqnarray}
It is identical to eq.(2.17). By taking $U_{-2}=U_{-1}=0$ in $\Lambda$ in
eq.(3.32), which are actually first-class constraints, (3.32) is
trivially reduced to the version of the first super KP Hamiltonian structure
obtained in ref.[10]. Therefore we have achieved the super KP bi-Hamiltonian
structure with eq.(3.22) as its associated superalgebra.

This bi-Hamiltonian structure (3.30) and (3.32) naturally leads to a set of
Lenard recursion relations connecting the conserved supercharges
$\int\Pi_{m}dX \equiv \int\Pi_{m}^{(2)}dX = $ $\int\Pi_{m-1}^{(1)}dX$:
\begin{eqnarray}
{\{ U_{i}, \int\Pi_{m+1}dX \}}_{1} = -{\{ U_{i}, \int\Pi_{m}dX \}}_{2}.
\end{eqnarray}
Futhermore, this set of conserved supercharges is in involution with respect
to the bi-Hamiltonian structure (3.30) and (3.32). For example,
\begin{eqnarray}
{\{ \int\Pi_{n}(X)dX, \int\Pi_{m}(Y)dY \}}_{2} &=&
\frac{1}{mn}\int SRes(K(\frac{\delta SRes\Lambda^{m}}{\delta\Lambda})
\frac{\delta SRes\Lambda^{n}}{\delta\Lambda})dX \nonumber\\
&=& \int SRes({[(\Lambda^{m})_{+},\Lambda]}\Lambda^{n-1})dX ~=~0.
\end{eqnarray}
The supercommutivity among $\int\Pi_{n}dX$'s under eq.(3.32) holds similarly
or due to the recursion relation (3.33). This proves the formal complete
integrability of the super KP hierarchy (2.6).

\vspace{30 pt}
\section{Super BKP Hierarchy}
\setcounter{equation}{0}
\vspace{5 pt}

Finally in this section, we construct the super BKP hierarchy --
a supersymmetric extension of the ordinary BKP hierarchy [19]. It is a system
of nonlinear super-differential evolution equations obtained from the super
KP hierarchy (2.6)
\begin{eqnarray}
\frac{\partial\Lambda}{\partial t_{m}} = {[(\Lambda^{m})_{+},\Lambda]}
\end{eqnarray}
with $m$ being odd positive integers and with the following anti-self-dual
constraints imposing on $\Lambda$:
\begin{eqnarray}
\Lambda = -\theta^{-1}\Lambda^{*}\theta.
\end{eqnarray}
Here the ``dual'' operation $*$ is defined to be
\begin{eqnarray}
\theta^{*} =-\theta, ~~~~~F(U_{i})^{*} =F(U_{i})
\end{eqnarray}
for any function $F(U_{i})$ and
\begin{eqnarray}
(PQ)^{*} = (-1)^{p(P)p(Q)}Q^{*}P^{*}
\end{eqnarray}
for any two pseudo-super-differential operators $P$ and $Q$. It follows that
for $n\geq 0$,
\begin{eqnarray}
\theta^{n*} = (-1)^{[(n+1)/2]}\theta^{n}, ~~~~~
\theta^{-n*} = (-1)^{[n/2]}\theta^{-n}
\end{eqnarray}
and for any $P$,
\begin{eqnarray}
SResP^{*} = SResP.
\end{eqnarray}
Now we see the constraints (4.2) is equivalent to letting $U_{i}$ satisfy
\begin{eqnarray}
U_{i} = \sum^{i}_{k=0}(-1)^{[k/2]+[(i-k)/2]} \left[ \begin{array}{c}
i+1\\k+1
\end{array} \right] U_{k}^{[i-k]}.
\end{eqnarray}
For the first a few of them, we have explictly
\begin{eqnarray}
& & U_{-2}=0,~~~U_{-1}=0,~~~U_{0}\equiv U_{0},~~~U_{1}\equiv U_{1}, \nonumber\\
& & U_{2}=\frac{1}{2}(U_{1}^{[1]}-U_{0}^{[2]}),~~~U_{3}=-U_{1}^{[2]},
\nonumber\\
& & U_{4}\equiv U_{4},~~~U_{5}\equiv U_{5},~~~\cdots,
\end{eqnarray}
and in general, half of the degrees of freedom of $U_{i}$ are eliminated by
eq.(4.7). Exactly, we have the following proposition: Eq.(4.2) holds if and
only if
\begin{eqnarray}
SRes(\Lambda^{2n+1}\theta^{-1})=0
\end{eqnarray}
and
\begin{eqnarray}
SRes\Lambda^{2n} =\frac{1}{2}SRes(\Lambda^{2n}\theta^{-1})^{[1]},
\end{eqnarray}
where $n=0,1,2,\ldots$ and $\Lambda^{0}\equiv U_{-2}\theta^{-1}$.

Proof: For necessity, by using eq.(4.6) we have
\begin{eqnarray}
SRes\Lambda^{m} &=& SRes\Lambda^{m*} ~=~ SRes(-\theta\Lambda\theta^{-1})^{m}
\nonumber\\
&=& (-1)^{m+1}SRes\Lambda^{m} +(-1)^{m}SRes(\Lambda^{m}\theta^{-1})^{[1]},
\nonumber
\end{eqnarray}
which leads to eq.(4.10) with even $m$; with $m$ being odd, we proceed
slightly differently: from eq(4.2)
\begin{eqnarray}
SRes(\Lambda^{m}\theta^{-1}) = -SRes(\theta^{-1}\Lambda^{m*}) \nonumber
\end{eqnarray}
and from eq.(4.6)
\begin{eqnarray}
SRes(\Lambda^{m}\theta^{-1}) = SRes(\Lambda^{m}\theta^{-1})^{*} =
SRes(\theta^{-1}\Lambda^{m*}), \nonumber
\end{eqnarray}
hence eq.(4.9) is true.

For sufficiency, we notice from the leading equations of (4.9) and (4.10)
that $U_{-2}=U_{-1}=0$, so we can always write
\begin{eqnarray}
B\equiv \Lambda +\theta^{-1}\Lambda^{*}\theta =b_{i}\theta^{-i} +
{\rm lower~order~terms}
\end{eqnarray}
where $i$ is certain positive integer. Because $B^{*}=\theta^{-1}B\theta$,
we find
\begin{eqnarray}
(-1)^{[i/2]}b_{i}\theta^{-i} = (-1)^{i}b_{i}\theta^{-i} +
{\rm lower~order~terms}, \nonumber
\end{eqnarray}
therefore $b_{i}$ vanish for $i=4n+1,4n+2~(n=0,1,2,\ldots)$ as identities.
Moreover, for $i=4n$,
\begin{eqnarray}
SRes(\Lambda^{2n+1}\theta^{-1}) &=& SRes((B-\theta^{-1}
\Lambda^{*}\theta)^{2n+1}\theta^{-1}) \nonumber\\
&=& (-1)^{2n+1}SRes((\theta^{-1}
\Lambda^{*}\theta)^{2n+1}\theta^{-1}) + (2n+1)b_{4n} \nonumber\\
&=& -SRes(\Lambda^{2n+1}\theta^{-1}) + (2n+1)b_{4n}, \nonumber
\end{eqnarray}
which yields $b_{4n}=0$ due to eq.(4.9). Similarly for $i=4n+3$,
\begin{eqnarray}
SRes\Lambda^{2n+2} &=& SRes(B-\theta^{-1}\Lambda^{*}\theta)^{2n+2}~=~
SRes(\theta^{-1}\Lambda^{*2n+2}\theta) +(2n+2)b_{4n+3} \nonumber\\
&=& SRes(\Lambda^{2n+2}\theta^{-1})^{[1]} -SRes\Lambda^{2n+2} +(2n+2)b_{4n+3},
\nonumber
\end{eqnarray}
which leads to $b_{4n+3}=0$ from eq.(4.10). Overall, eq.(4.11) is equal to
zero as desired. (QED)

Now we consider the consistency of the super BKP hierarchy, that is obtained
via our final proposition: Eq.(4.1) is compatible with eq.(4.2).

Proof: We need to show
\begin{eqnarray}
\frac{\partial(\Lambda+\theta^{-1}\Lambda^{*}\theta)}{\partial t_{m}} = 0,~~~~~
m=1,3,5,\ldots
\end{eqnarray}
which is equivalent to
\begin{eqnarray}
\theta^{-1}{[(\Lambda^{m})_{+},\Lambda]}^{*}\theta = -
{[(\Lambda^{m})_{+},\Lambda]}
\end{eqnarray}
with $\Lambda+\theta^{-1}\Lambda^{*}\theta =0$. Indeed,
\begin{eqnarray}
{[(\Lambda^{m})_{+},\Lambda]}^{*} &=& \Lambda^{*}(\Lambda^{m})_{+}^{*}
-(\Lambda^{m})_{+}^{*}\Lambda^{*} \nonumber\\
&=& \theta\Lambda\theta^{-1}(\theta\Lambda^{m}\theta^{-1})_{+}
-(\theta\Lambda^{m}\theta^{-1})_{+}\theta\Lambda\theta^{-1} \nonumber\\
&=& \theta\Lambda\theta^{-1}(\theta(\Lambda^{m})_{+}\theta^{-1}
-SRes(\Lambda^{m}\theta^{-1})^{[1]}\theta^{-1}) \nonumber\\
& & -(\theta(\Lambda^{m})_{+}
\theta^{-1} -SRes(\Lambda^{m}\theta^{-1})^{[1]}\theta^{-1})
\theta\Lambda\theta^{-1} \nonumber\\
&=& -\theta{[(\Lambda^{m})_{+},\Lambda]}\theta^{-1} \nonumber
\end{eqnarray}
where we have used eq.(4.9). (QED)

In conclusion we note through the natural reduction (4.2) the super KP
bi-Hamiltonian structure obtained in earlier sections is expected to give
rise to its super BKP counterpart. We conjecture the
constraints (4.2) are of first class so that this Hamiltonian
reduction can be accomplished without Dirac's prescription.

\vspace{30 pt}
Acknowledgement: I would like to thank Y.-S. Wu for conversations on many
related subjects.
This work is supported in part by the Graduate Research Fellowship of
University of Utah.

\vspace{30 pt}
\begin{center}
{\large REFERENCES}
\end{center}
\begin{itemize}
\vspace{5 pt}

\item[1.] For review, see
I. M. Gelfand and L. A. Dickey, Russ. Math. Surv. 30 (1975) 77;
Funct. Anal. Appl. 10 (1976) 259; I. M. Gelfand and I. Dorfman, Funct.
Anal. Appl. 15 (1981) 173; B. A. Kupershmidt and G. Wilson, Invent. Math.
62 (1981) 403; V. Drinfel'd and V. Sokolov, Journ. Sov. Math. 30 (1985) 1975.
\item[2.] For review, see
M. Sato, RIMS Kokyuroku 439 (1981) 30; E. Date, M. Jimbo, M.
Kashiwara and T. Miwa, in Proc. of RIMS Symposium on Nonlinear Integrable
Systems, eds. M. Jimbo and T. Miwa, (World Scientific, Singapore, 1983);
G. Segal and G. Wilson, Publ. IHES 61 (1985) 1.
\item[3.] M. Adler, Invent. Math. 50 (1979) 219.
\item[4.] I. M. Gelfand and L. A. Dickey, Russ. Math. Surv. 30 (1975) 77.
\item[5.] Y. Watanabe, Ann. di Mat. Pura Appl. 86 (1984) 77.
\item[6.] L. A. Dickey, Annals New York Academy of Sciences, (1987) 131.
\item[7.] Yu. I. Manin and A. O. Radul, Comm. Math. Phys. 98 (1985) 65.
\item[8.] J. M. Figueroa-O'Farrill, J. Mas and E. Ramos, Preprint
KUL-TF-91/17; KUL-TF-91/19.
\item[9.] Y. Watanabe, Lett. Math. Phys. 14 (1987) 263.
\item[10.] F. Yu, University of Utah preprint UU-HEP-91/12.
\item[11.] A. A. Belavin, Adv. Study Pure Math. 19 (1989) 117;
V. A. Fateev and S. L. Lykyanov, Int. J. Mod. Phys. A3
(1988) 507; I. Bakas, Phys. Lett.
B213 (1988) 313; B219 (1989) 283; P. Di Francesco,
C. Itzykson and J.-B. Zuber, Saclay/Princeton
preprint SPhT/90-149 and PUPT-1211.
\item[12.] A. B. Zamolodchikov, Theor. Math. Phys. 65 (1985) 1205;
A. B. Zamolodchikov and V. A. Fateev, Nucl. Phys. B280 [FS18] (1987) 644.
\item[13.] F. Yu and Y.-S. Wu, Phys. Lett. 263B (1991) 220;
K. Yamagishi, Phys. Lett. 259B (1991) 436.
\item[14.] C. N. Pope, L. Romans and X. Shen, Phys. Lett. 236B (1990) 173;
Phys. Lett. 242B (1990) 401; Phys. Lett. 245B (1990) 72.
\item[15.]
I. Bakas, Phys. Lett. 228B (1989) 57; I. Bakas and E. B. Kiritsis,
LBL-29394, UCB-PTH-90/33, UMD-PP90-272.
\item[16.] F. Yu ana Y.-S. Wu, University of Utah preprint UU-HEP-91/09.
\item[17.] E. Bergshoeff, C. N. Pope, L. J. Romans, E. Sezgin and X. Shen,
Phys. Lett. 245B (1990) 447.
\item[18.] J. M. Figueroa-O'Farrill and E. Ramos, Preprint
KUL-TF-91/6; KUL-TF-91/13; KUL-TF-91/14.
\item[19] E. Date, M. Jimbo, M.
Kashiwara and T. Miwa, in Proc. of RIMS Symposium on Nonlinear Integrable
Systems, eds. M. Jimbo and T. Miwa, (World Scientific, Singapore, 1983).
\item[20.] D. Gross and A. Migdal, Phys. Rev. Lett. 64 (1990) 127; M. Douglas
and S. Shenker, Nucl. Phys. B335 (1990) 635; E. Brezin and V. Kazakov, Phys.
Lett. 236B (1990) 144.
\item[21.] M. Douglas, Phys. Lett. 238B (1990) 176; T. Banks, M. Douglas,
N. Seiberg and S. Shenker, Phys. Lett. 238B (1990) 279;
D. Gross and A. Migdal, Nucl. Phys. B340 (1990) 333;
P. Di Francesco and D. Kutasov, Nucl. Phys. B342 (1990) 589.
\item[22.] E. Verlinde and H. Verlinde, Nucl. Phys. B348 (1991) 457;
M. Fukuma, H. Kawai and R. Nakayama, Tokyo/KEK preprint UT-562,
KEK-TH-251 (1990); J. Goeree, Utrecht preprint THU-19 (1990).

\end{itemize}

\end{document}